\documentstyle[prl,aps,multicol]{revtex}
\begin{document}
\draft
\widetext

\title{Disordered Critical Wave functions
 in  Random Bond Models in Two Dimensions
\\ --- Random Lattice  Fermions at $E=0$  without Doubling ---}
\author{Yasuhiro Hatsugai$^{1*}$, Xiao-Gang Wen$^2$, and Mahito Kohmoto$^3$}
\address{$^1$ Department of Applied Physics,
 University of Tokyo,
 7-3-1 Hongo Bunkyo-ku, Tokyo 113, Japan}
\address{$^2$ Department of Physics, Massachusetts Institute of
Technology,  Cambridge, Massachusetts 02139}
\address{$^3$ Institute for Solid State Physics,
 University of Tokyo,
 7-22-1 Roppongi Minato-ku
 Tokyo 106, Japan}
\date{March 14, 1996}
\maketitle

\widetext
\begin{abstract}
\leftskip 54.8pt
\rightskip 54.8pt

Random bond Hamiltonians of the $\pi$ flux state
on the square lattice are investigated.
It has a special symmetry
and all states are paired except the ones with zero energy.
Because of this, there are always zero-energy states (or zero-modes).
The states near the zero-modes are described by
massless Dirac fermions.
For the zero-mode
in a system with odd numbers of sites,
we can construct a random
lattice fermion without a doubling
and
quite large systems ( up to $801 \times 801$) are
treated numerically.
 We clearly demonstrate that
the zero-mode is given by a critical wave function.
For the multifractal behavior of the wave function,
the agreement with the effective field theory is  demonstrated.

\end{abstract}


 \begin{multicols}{2}

\narrowtext

Random critical points in {\it two dimensions} appear in quantum
Hall systems and in systems with spin-orbit coupling.
Both critical points have important experimental implications, and
both are in strong coupling limit. Because of that we are still
unable to calculate their critical exponents analytically,
despite the efforts of  more than a decade.
Recently, a class of exactly solvable two-dimensional random
critical points was found.\cite{lfsg,cmw1}
Those critical points appear
in models of Dirac fermions coupled to random gauge fields.
However, it was pointed out that those critical points
contain infinite number of relevant directions. Thus it seems
that it is almost impossible to realize those random critical points
in a lattice model.

In this paper, we study  random bond models on the square lattice
with only nearest neighbor hopping.
 The models
 have $\pi$ flux per plaquette on the average.
When all hopping matrix elements are real (in this case the model
will be called random hopping (RH) model),
we find that with weak randomness and in a
continuum approximation, the states
near zero energy are described by two copies of Dirac fermions
coupled to {\it imaginary} random chemical potential, real
random mass, and {\it imaginary} gauge fields.
The results in Ref.\cite{lfsg,bernard},
 suggest that the continuum model
has a line of random critical points, and
due to the {\it imaginary} gauge fields and the
{\it imaginary} chemical potential, the critical
points has only one marginally relevant direction if
we fix the energy to be zero.
Thus it could be possible
to fine tune   parameters   to reach
the  critical line.
In fact, the zero-energy states of the
fine tuned model are found to have
a multifractal behavior and are critical.\cite{halsey}
We also present direct
numerical results indicating that the zero-energy states
in the generic random bond model also appear to be critical with
multifractal behaviors at least for the systems of sizes
up to $801\times 801$.
The multifractal scaling function $f(\alpha)$
for the fine-tuned RH model
is calculated numerically which agrees with the
exact result of the  continuum theory.\cite{lfsg,cmw1}.
Another interesting point in our treatment is that
we can avoid the fermion doubling as far as the zero-energy states are
concerned.

 In one dimension, some of the wave functions of the quasiperiodic
systems show the critical behavior clearly.\cite{hk}
In two dimensions,
it was difficult to show the criticality clearly
 due to the limitation of available system sizes.
However, since our models have several specialties as described below,
we can treat quite large two dimensional systems up to $801\times 801$
which enables a clear demonstration.

The wave functions in
a two-dimensional random system are believed to be
always localized
 if the system has the time reversal symmetry.
Thus it is quite interesting that our  models
have random critical points.
There are several reasons why the critical states are
allowed in our models
 in two dimensions.
Our Random bond  model has a very special property that
the random Hamiltonian anti-commute with the operator
$\gamma$ defined below.
This means that the eigenstates always appear in pairs
with energies $E$ and $ -E$.
Therefore the states near zero energy
are quite special, and one may wonder if they are critical states
or not.
We may also consider
 a pure system with a small diagonal hopping
$m$ (which break the symmetry
mentioned above).\cite{hkw}
The  symmetry is recovered by taking a limit $m\to 0$.
When $m \neq 0$, there is an energy gap near $E=0$
and there are
two energy bands.
The Hall conductance of the lower band is $+{\rm sign}(m)$
and
$-{\rm sign}(m)$ for the higher one.
Since the nonzero Hall conductance is carried by extended states,
there are paired extended states in the lower and the higher bands.
By taking   $m\to 0$,
the paired extended states merges and disappear
at $E=0$.
It suggests the criticality
of the $E=0$ states.
Also from a point of the perturbation theory,
the density
of states of the pure state vanishes at zero energy.
Therefore a usual treatment of the purtabative consideration
has to be modified.  This property, we believe, is also
responsible for the appearance of the critical states at zero energy.

Our Hamiltonian is
\begin{equation}
H=\sum_{<i,j>} c_i^\dagger t_{ij} c_j+h.c.
\end{equation}
where the summation is over the nearest-neighbor bonds.
We shall analyze the following two types of
random bond models  near the $\pi$ flux:
(i) {\it  random gauge model}
$t_{j+\hat x,j}=(-)^{j_y}e^{i\delta\theta_x(j)}$,
$t_{j+\hat y,j}=         e^{i\delta\theta_y(j)}$,
(ii) {\it random hopping (RH) model},
$t_{j+\hat x,j}=(-)^{j_y}+\delta t_x$,
$t_{j+\hat y,j}=  1 +  \delta t_y$.
Here $ \delta\theta_{x,y}(j) $
and $\delta t_{x,y}$
are  real random variables.
 Numerical results of the wave functions
for these two models are qualitatively similar.
 The model (i) belongs to the unitary
ensembles.
The model (ii) preserves the time reversal symmetry and
 belongs to the orthogonal ensembles
which
is suitable
for a comparison with the critical continuum theory.

In the absence of the randomness,
the system is invariant under the translation and
the Hamiltonian
is written in the momentum space:
\begin{equation}
H_0=
\int_{-\pi}^{\pi}{\frac {dk_x}{2\pi}}
\int_0^{\pi}{\frac {dk_y}{2\pi}}
 \psi^\dagger_k \pmatrix{\cos k_y & \cos k_x  \cr
                                     \cos k_x &-\cos k_y  \cr}
\psi_k
\label{H0}
\end{equation}
with
$\psi^\dagger_k= ( c^\dagger_k, c^\dagger_{k+(0,\pi)})$.
We see that the zero-energy state is formed by the fermions with
momentum $\mbox{\boldmath $k$}_1=( \pi/2, \pi/2)$ and
$\mbox{\boldmath $k$}_2=( - \pi/2, \pi/2)$.
Expanding it near $\mbox{\boldmath $k$}_1$ and
$\mbox{\boldmath $k$}_2$, we find
in the continuum approximation, $H_0$ becomes
\begin{equation}
H_0=2i\int d^2x \Psi^\dagger \left[
\pmatrix{\sigma_1&0\cr 0& -\sigma_1\cr} \partial_x +
\pmatrix{\sigma_3&0\cr 0&  \sigma_3\cr} \partial_y  \right]\Psi
\end{equation}
with
$
 \Psi^\dagger  =
1/(4\pi^2)\int  {d \delta k_x\delta k_y}
e^{-i\delta k_x x-i\delta k_y y }
(\psi^\dagger _{k_1}, \psi^\dagger_{k_2})
$
and
$
c_{j_x,j_y}
=
(i^{j_x + j_y}, i^{j_x - j_y}, i^{-j_x + j_y}, i^{-j_x - j_y} )\Psi$
where $\sigma$'s are the Pauli matrices.
We assume that $\Psi$  take the same value
on four sites $(x,y)=(2m,2n)$, $(2m+1,2n)$, $(2m,2n+1)$, $(2m+1,2n+1)$.
Note that for weak impurities, the states near zero energy are described by
the smooth function   $\Psi(x)$.
Therefore we can derive a continuum theory
to describe those states near zero energy.

Before deriving the continuum theory with randomness,
we  first study the symmetries of the lattice Hamiltonian.
Since the square lattice is bipartite,  the transformation
 $c_{j_x,j_y}\to c'_{j_x,j_y}=(-1)^{j_x+j_y}c_{j_x,j_y}$ induces
sign change of the Hamiltonian.
It is expressed as
$
H(x)  \to   \gamma^\dagger  H(x) \gamma = - H(x)
$,
$\gamma = \sigma_1\otimes \sigma_1$
where
$
H= \int d^2 x \Psi(x) ^ \dagger H(x) \Psi(x)
$
is the continuum Hamiltonian.
The contribution from the randomness $\delta H\equiv H-H_0$ has a form
$
\delta H=\int d^2x \Psi^\dagger g(x) \Psi
$
if we ignore the terms containing derivatives.
The above  condition implies that
the most general form of the $g(x)$ is given by
a linear combination of the
4 by 4 matrices:
$
\gamma_1=\sigma_2\otimes I
$,
$
\gamma_2=\sigma_1\otimes \sigma_2
$,
$
\gamma_3=-\sigma_2\otimes\sigma_1
$, 
$
\gamma_4=I  \otimes \sigma_2
$
with  coefficients $2a_i(x)$, $i=1,...,4$.
The total Hamiltonian, after a change of basis to make
$\gamma=-\sigma_3\otimes I $, ( $ \Psi=T\Psi'$,
$T=(I \otimes I +i\sigma_2\otimes\sigma_1)/\sqrt{2}$), is given by
\begin{eqnarray}
H &=& \int d^2x \Psi'^\dagger
2  ( I  \otimes  \sigma_1 )
\pmatrix{0&D_+\cr D_-&0\cr} \Psi', \\
\label{Hc}
D_\pm &=& i[(\partial_x\mp a_1)\sigma_1\mp (\partial_y \mp a_2)\sigma_3]
\pm i a_3  \mp \sigma_2 a_4.
\end{eqnarray}
Note that the averaged Hamiltonian is invariant under translation
$(x,y)\to (x+1,y)$. This requires that $\langle a_i\rangle =0$ for $i=1,2,3$.
The averaged Hamiltonian is also invariant under reflection
$(x,y)\to (x,-y)$. This implies that $\langle a_4\rangle=0$.

The zero-energy state of $H$
satisfies
 \begin{equation}
 D_+\psi_+=0,\ \ \  D_-\psi_-=0,
 \label{eq:twodirac}
 \end{equation}
where $\pmatrix{\psi_-\cr \psi_+\cr}=\Psi'$.
Thus the random bond model
 model  contains two Dirac fermions in a continuum approximation,
which is the famous fermion doubling of the lattice fermions.
It causes several difficulties in numerical calculations.
However, as we discuss below,
{\em we can avoid the doubling as far as
the $E=0$ state is concerned}.

Now let us  discuss
about the doubling which arises from the lattice models.
Since our model is
a nearest neighbor hopping model
on a bipartite lattice,
we can write the Hamiltonian  as
\begin{equation}%
H= (\{c^\dagger_+\},\{c^\dagger_-\})
 \pmatrix{
O&{\cal D}\cr
{\cal D}^\dagger&O\cr
}
\pmatrix{
\{c_+\}\cr
\{c_-\}\cr
}
\end{equation}%
where
$\{c_+\}=\{c_{j_x,j_y}|
j_x+j_y {\rm \ is\ even\ } \}=\{c_{++},c_{--}\}$ is a set of
fermion operators
at one of the sublattices ($+$) and
$\{c_-\}=\{c_{j_x,j_y}|
j_x+j_y {\rm \ is\ odd\ } \}=\{c_{+-},c_{-+}\}$ is the other
($-$).
Here let us assume a system  to be
$L_x\times L_y$ with both odd $L_x$ and $L_y$
with a fixed boundary condition.
( We label the sites as $(1,1), \cdots, (L_x,L_y)$.)
Then a total number of sites $L_x L_y=N_+ +N_- $ is odd with
$N_+=N_-+1$ where $N_+=\#\{+ {\rm \ sites}\}$,
  $N_-=\#\{- {\rm\ sites}\}$.
Therefore ${\cal D}$ is a
$N_+\times N_-=N_+\times (N_+-1)$ rectangular matrix.
The Schr\"{o}dinger equation
is reduced to
${\cal D} {\cal D}^\dagger \phi_+= E^2 \phi_+ $ and
${\cal D}^\dagger {\cal D} \phi_- = E^2 \phi_-  $.
Any state with $E\neq 0$ has a pair state at $-E$
as
$^t(
\phi_+,
\phi_-)_{-E}
=^t(
\phi_+,
-\phi_-)_{E}$.
Since $\det  \pmatrix{
O&{\cal D}\cr
{\cal D}^\dagger&O\cr
}= 0$ ( note that ${\cal D}$ is a form of $N_+\times N_+-1$),
there is always zero-mode.
The wave function at $E=0$ satisfies
\begin{equation}%
{\cal D}^\dagger \phi^{E=0}_+= 0,
\ \
{\cal D} \phi^{E=0}_- = 0.
\end{equation}%
Since $\dim \phi^{E=0}_+ =N_+$, the first one gives
$N_-=N_+-1$ equations for
$N_+$ variables and
there is a {\em one } dimensional non vanishing solution.
(Therefore this zero-mode is not degenerate in general.)
On the other hand,
since $\dim \phi^{E=0}_- =N_-=N_+-1$, the second one gives
$N_+$ equations for
$N_-=N_+-1$ variables, this is overdetermined.
Therefore we have
$\phi^{E=0}_- = 0$.

It gives an important restriction in the continuum model.
It implies
 \begin{equation}%
 \psi_{-} = 0
 \end{equation}%
in (\ref{eq:twodirac})
since the transformation matrix between $|\Psi\rangle'$'s and the site
operators' $U T^{-1} U^{-1}$, only mixes operators within each sublattices.
Therefore there is no  degeneracy ( doubling )
for the $E=0$ state.
This $E=0$ state is described by  a usual two
component spinor.
Therefore  the independent component is not four but two
as far as  the $E=0$ state is concerned.
We could avoid the doubling by using the above special  conditions.

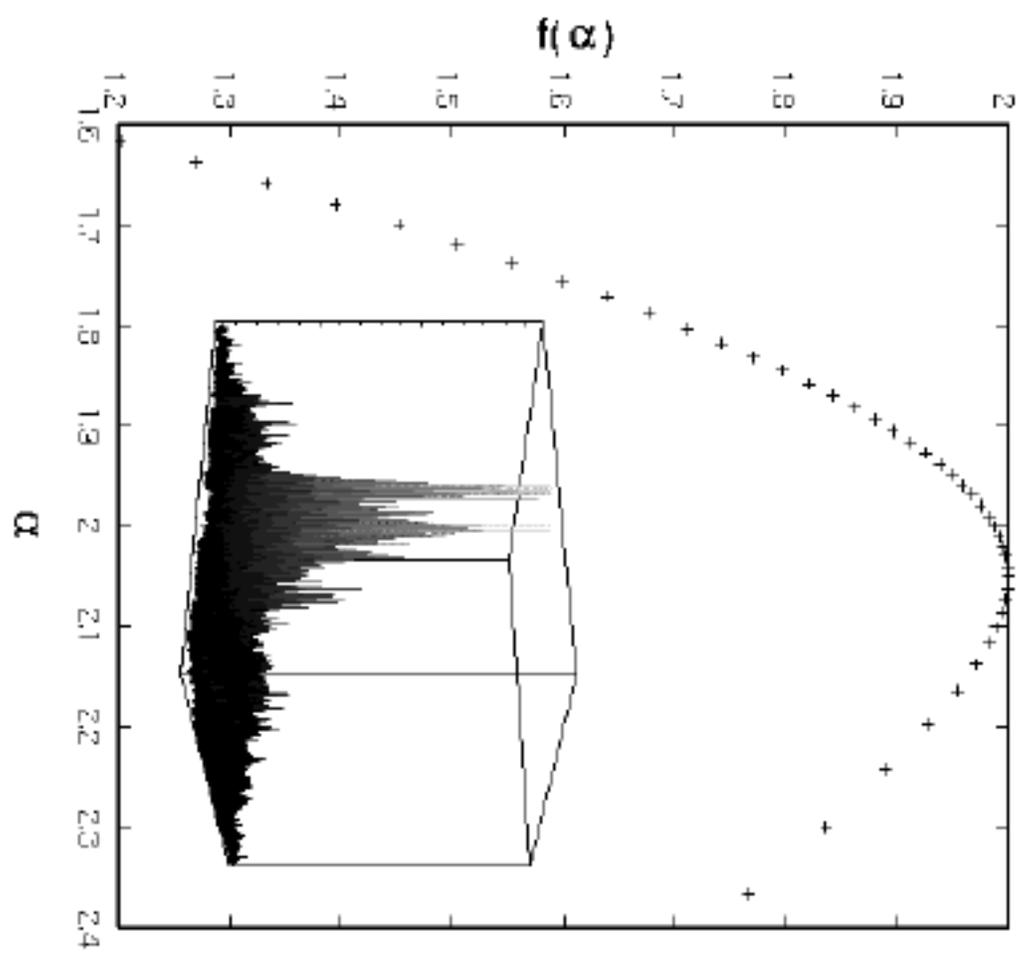
\begin{figure}
\caption{
The zero-mode wave function of the random gauge model
and the corresponding $f(\alpha)$.
The system size is
$801\times 801$  and   $V=0.2 $.
The errors by the finite size corrections in the box sizes
is about the symbol sizes  within
the shown region but is larger in the other.
 \label{fig:wavefunc}}
\end{figure}

{\it Random Gauge Model}. ---
Now let us show numerical results
for the random gauge model.
In the previous  numerical calculations for
the $\pi$ flux state with randomness\cite{hl},
it was not possible to treat
large two-dimensional systems.
In the present work, however, we
only need the  eigenstate of a semi positive definite
operator ${\cal D} {\cal D}\dagger$ with zero eigen value.
This and the fact that the $E=0$ eigen states are generally
not degenerate enable
us to treat large size systems up to $801\times 801$.
We take $ \delta\theta_x(i) $ and $ \delta\theta_y(i) $
to be uniform random numbers between $-V/2$ to $V/2$.
In  Fig.~\ref{fig:wavefunc}, the wave function is shown for $V=0.2$
with $801\times 801$ system where it is coarse grained over
4 sites on the plaquette.
It seems to be neither localized nor extended
in a usual manner.
To understand the nature of the wave functions,
the function $f(\alpha) $ obtained numerically are also shown.
The box size dependence
is carefully included in the calculations.\cite{hk}
The maximum of $f(\alpha)$ gives a Hausdorf dimension
which is always two here since the wave functions are on the square lattice.
The value $\alpha_0$ which gives the maximum of $f(\alpha)$
gives scaling of the dominant parts of the wave function.
It is two for an extended states which gives uniform nonsingular scaling.
For a critical state, $\alpha_0 \neq 2 $ and has
 a singular scaling.\cite{huc,ahk}
In our case, it apparently deviates from $2$ which means
the zero-mode is  critical.

{\it Random Hopping Model}. ---
The properties of random Dirac fermion models have been studied in detail
lately.
We study the random hopping model to make a comparison with the
continuum theory.
In the continuum approximation,
the fact that $\delta t_{x,y}$'s are real further requires
$\gamma g(x) \gamma =g(x)^*$.
Thus we take with the coefficients $a_i(x)$, $i=1,...,4$ to be real.
Note that $D_\pm$ happens to be the Hamiltonian of  Dirac fermions
with {\it imaginary} random gauge potentials $(a_1,a_2)$,
an {\it imaginary} random chemical potential $a_4$,
and a real random mass $a_3$.
Let $(a_1,a_2)$, $2a_3$ and $2a_4$ to have a Gaussian distribution
with width $g_A$, $g_V$ and $g_M$ respectively. Averaging over
randomness generates four-fermion interactions with
$(g_A, g_V, g_M)$ as coupling constants. The one-loop
beta functions for the above coupling constants is calculated in
Ref.\cite{bernard} as
 \begin{eqnarray}%
   \dot g_A &=& 32g_M g_V \\
  \dot g_V+ \dot g_M &=& -8(g_V-g_M)^2 \\
  \dot g_V- \dot g_M &=& -8(g_V+g_M)(g_V-g_M)-8g_A(g_V-g_M)
  \end{eqnarray}%
(Note comparing to Ref.\cite{bernard}, here we have
an {\it imaginary} random chemical potential, a real random mass
and an {\it imaginary} random gauge potential.
Thus the beta functions are modified accordingly with
$(g_A, g_V, g_M) \to (-g_A, -g_V, g_M)$.)
The renormalization-group (RG) equation has a fixed line
$(g_A, g_V, g_M) = (g_A, 0, 0)$.
Around the fixed line,
$g_V-g_M$ is irrelevent (note $g_A>0$)
and thus we may set $g_V=g_M$.
 In this case
$g_V+g_M$ does not flow, and  non-zero $g_V+g_M$ will drive $g_A \to \infty$.
Thus the fixed line has at least one (marginally) relevent direction.
Two-fermion operators are also relevent. But
those two-fermion operators do not appear since $<a_i>=0$, $i=1,...,4$
due to the discrete symmetries.
Because of
the {\it imaginary} random gauge potential,
the results in Refs.\cite{cmw}, \cite{bernard}
imply that all charge neutral operators
are irrelevant, except two-fermion operators and four-fermion operators
discussed above.
Thus the fixed line has only one marginally relevent direction.
It suggests that the RH model have a line of critical
points at zero energy which is described by the line of
critical points in the random Dirac fermion model. Those results
further suggest that,
once we fix the energy to be zero, there
is only one  marginally relevant direction for the critical line.
Thus the critical line in the RH model contains
two relevant directions.
Due to its relation to the random Dirac fermion model,
the critical line in the RH model is exactly
solvable, and critical exponents can be calculated exactly.

To describe the critical line, set $a_3=a_4=0$.
Let us assume the probability distribution of
$(a_1({\bf r}),a_2({\bf r}))$
is given by
$
P\propto {\rm exp}(-\int d^2r \frac{1}{2g_A}[a_1^2({\bf r}) + a_2^2({\bf r})])
$
where $g_A$, characterizing the strength of the gauge field fluctuations, is
exactly marginal. Thus $g_A$ can be used to parameterize different
critical points on the critical line.
In general critical exponents are functions of $g_A$.

Note that the energy $E$ couples to an operator $\Psi_+^\dagger \Psi_+$.
Since $\Psi_+$ and $\Psi_+^\dagger$ belong to two independent
Dirac fermion models described by $D_+$ and $D_-$ respectively,
The scaling dimension of $\Psi_+^\dagger \Psi_+$ is simply the sum
of the scaling dimensions of $\Psi_+$ and $\Psi_+^\dagger$.
 From Ref.\cite{cmw} , we see that
the scaling dimensions of $\Psi_+$ and $\Psi_+^\dagger$ are equal to 1/2,
which implies that
the scaling dimension of $\Psi_+^\dagger \Psi_+$ is 1.
This means that the dynamics exponent $z=1$ and the density
of states scales as $N(E)\propto |E|$; both exponents
are independent of $g_A$.

In the continuum theory  (\ref{Hc}), the zero-energy state can be solved
exactly \cite{lfsg}. Introducing $\phi$ and $\chi$ through
$a_\mu=\partial_\mu \phi+\epsilon_{\mu\nu}\partial_\nu \chi$.
One can show that the zero-energy state (with, for example, $\gamma=1$)
is given by
\begin{equation}
\Psi_+\propto e^{i\phi+\chi}
\label{wf}
\end{equation}
 and the probability distribution of
$\phi$ and $\chi$ is given by
$P\propto {\rm exp}(-\int d^2r \frac{1}{2g_A}
[(\partial \phi({\bf r}) )^2+
[(\partial \chi({\bf r}) )^2]).
$
The wave function (\ref{wf}) with this distribution
was studied in detail in Ref.\cite{lfsg} \cite{cmw1},
and was shown to have
a multifractal structure.
The exact multifractal scaling function $f(\alpha)$
is defined on $d^{\ }_-\leq\alpha \leq d^{\ }_+$ and
is given by\cite{cmw1}
\begin{equation}
f(\alpha)=8\frac{ (d_+-\alpha)(\alpha-d_-) }{(d_+ -d_-)^2}
\label{eq:fa}
\end{equation}
where
$
d^{\ }_{\pm}=2(1\pm \sqrt{\frac{g_A}{2\pi}} )^2
$
for $g_A< 2\pi$ and
$
d^{\ }_{+}=8\sqrt{\frac{g_A}{2\pi}},\ \ \ d_-=0
$
for $g_A> 2\pi$.
We see that for weak randomness ($g_A<2\pi$) $f(\alpha)$
is peaked at
\begin{equation}
\alpha_{0}=2+{g_A\over \pi}
\label{eq:amax}
\end{equation}

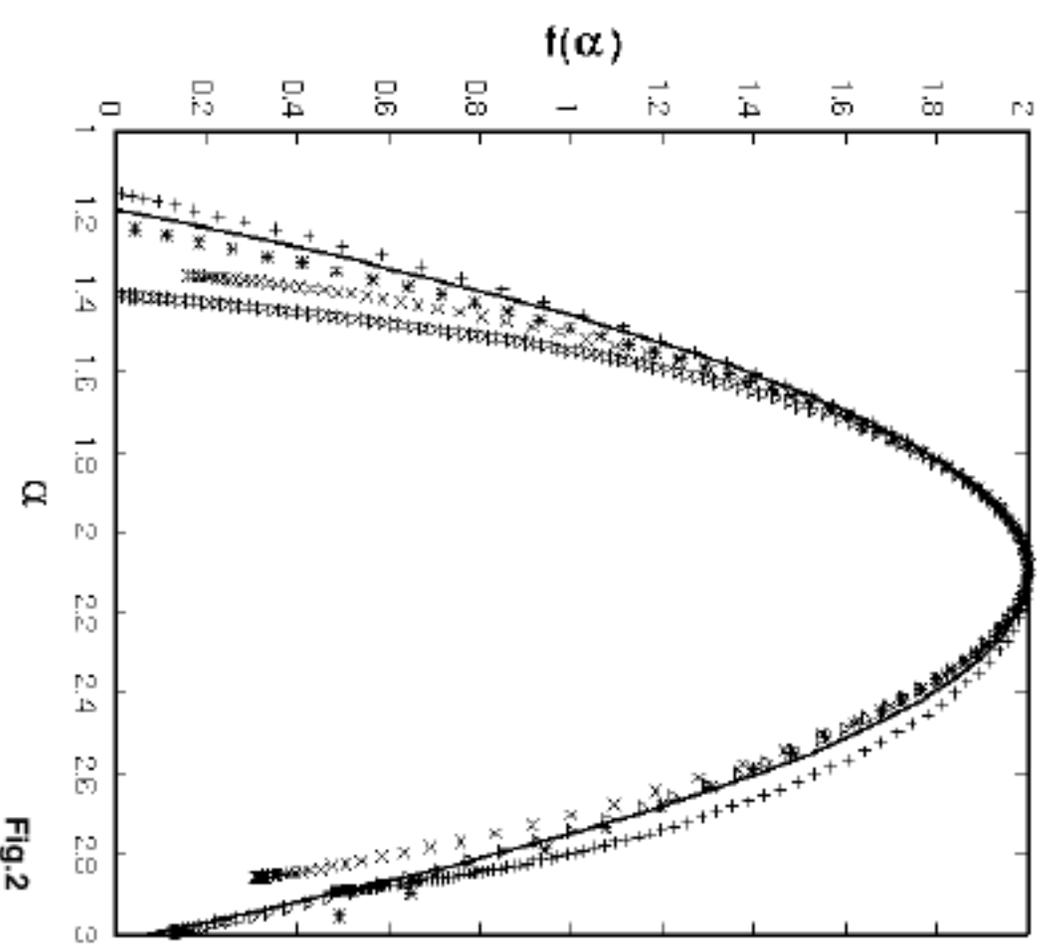
\begin{figure}
\caption{
Functions $f(\alpha)$'s for the zero-mode wave functions
of the RH model:
$ 401\times 401$ and
$V=0.3$.
The solid line is the analytical result. The different
symbols are for  the different randomness realizations.
 \label{fig:amax}}
\end{figure}

The parameters
$a_i({\bf r})$ in the continuum model is directly related to the lattice
randomness. We find that
$
a_1(j) = -\Lambda \delta t_{j,j+\hat x} (-)^{j_x}
$,
$
a_2(j) =-\Lambda \delta t_{j,j+\hat y} (-)^{j_x+j_y}
$,
$
a_3(j) = \Lambda \delta t_{j,j+\hat x} (-)^{j_x+j_y}
$ and
$
a_4(j)=-\Lambda \delta t_{j,j+\hat y} (-)^{j_y}
$
with a dimensional parameter $\Lambda $.
Therefore if we chose $\delta t_{ij}$ to satisfy
$\delta t_{j,j+\hat x}= \delta t_{j+\hat y,j+\hat y+\hat x}$
and
$\delta t_{j,j+\hat y} =-\delta t_{j+\hat x,j+\hat x+\hat y} $.
Then $g_V=g_M=0$ and the lattice model will be on the critical line.
We used this parametrization and performed a numerical calculation on the
above fine tuned RH model.
We set $\delta t_{x,y}$
as Gaussian random numbers with a variance $V^2$ ($g_A=\Lambda ^2 V^2$).
In   Fig.~\ref{fig:amax}, we have plotted
a singular scaling function
 $f(\alpha) $.
In the calculation, $g_A$ is obtained from a value of $\alpha_0$.
The $f(\alpha)$ obtained from numerical calculation
agrees well with the exact theoretical result Eq.~(\ref{eq:fa}).

In summary, we find that
in the continuum approximation,
the RH model at zero energy  has a critical
line with only one marginally relevent direction. This analytical
result is  supported by numerical calculations on the fine-tuned
 RH model.
Therefore the RH model is described by the continuum random Dirac fermions
as far as critical behavior is concerned.
We also find that the zero-energy states of
the general random bond models also appear to be critical
according to our numerical
results.
However, the numerical calculations can not eliminate a possibility
of localization with huge correlation length ( more than $800$ lattice
spacings)
which may appear in our problem due to the marginally relevent direction.

This work was supported in part by
Grant-in-Aid
from the Ministry of Education, Science and Culture
of Japan (YH, XGW, MK) and
NSF grants
DMR-9411574 (XGW).
XGW also acknowledges the support from
A.P. Sloan Foundation.

 \end{multicols}

\end{document}